\documentclass[fleqn,10pt]{wlscirep}
\usepackage[utf8]{inputenc}
\usepackage[T1]{fontenc}
\usepackage{hyperref}
\usepackage{graphics}
\usepackage{epsfig}
\usepackage{xspace}
\usepackage{wrapfig,enumitem}
\usepackage{multirow}
\usepackage{subcaption}

\title{A multi-centre polyp detection and segmentation dataset for generalisability assessment}
\author[1,2,3,*]{Sharib Ali}
\author[5,6,18]{Debesh Jha}
\author[7]{Noha Ghatwary}
\author[8,$\dag$]{Stefano {Realdon}}
\author[10,$\dag$]{Renato {Cannizzaro}}
\author[12]{Osama E. Salem}
\author[11]{Dominique {Lamarque}}
\author[13]{Christian Daul}
\author[5,6]{Michael A. Riegler}
\author[17]{Kim V. {A}nonsen}
\author[15]{Andreas Petlund} 
\author[5,16]{{P{\aa}l} Halvorsen}
\author[1,2]{Jens Rittscher}
\author[9,14,15$\dag$]{Thomas de Lange}
\author[2,4,$\dag$]{James E. East}

\affil[1]{School of Computing, University of Leeds, LS2 9JT, Leeds, United Kingdom}
\affil[2]{Institute of Biomedical Engineering, Department of Engineering Science, University of Oxford, OX3 7DQ, Oxford, United Kingdom}
\affil[3]{Oxford National Institute for Health Research Biomedical Research centre, OX4 2PG, Oxford, United Kingdom}
\affil[4]{Translational Gastroenterology Unit, Experimental Medicine Div., John Radcliffe Hospital, University of Oxford, OX3 9DU, Oxford, United Kingdom}
\affil[5]{SimulaMet, Pilestredet 52, 0167 Oslo, Norway}
\affil[6]{Department of Computer Science, UiT The Arctic University of Norway, Hansine Hansens veg 18, 9019 Tromsø, Norway}
\affil[7]{Computer Engineering Department, Arab Academy for Science and Technology,Smart Village, Giza, Egypt}
\affil[8]{Veneto Institute of Oncology IOV-IRCCS, Via Gattamelata, 64, 35128 Padua, Italy}
\affil[9]{Medical Department, Sahlgrenska University Hospital-Mölndal, Blå stråket 5, 413 45 Göteborg, Sweden}
\affil[10]{CRO Centro Riferimento Oncologico IRCCS Aviano Italy, Via Franco Gallini, 2, 33081 Aviano PN, Italy}
\affil[11]{Universit{\'e} de Versailles St-Quentin en Yvelines, H{\^o}pital Ambroise Par{\'e}, 9 Av. Charles de Gaulle, 92100 Boulogne-Billancourt, France}
\affil[12]{Faculty of Medicine, University of Alexandria, 21131, Alexandria, Egypt}
\affil[13]{CRAN UMR 7039, Universit\'e de Lorraine and CNRS, F-54010, Vand{\oe}uvre-L{\`es}-Nancy, France}
\affil[14]{Department of Molecular and Clinical Medicine, Sahlgrenska Academy, University of Gothenburg, 41345 Göteborg, Sweden}
\affil[15]{Augere Medical, Nedre Vaskegang 6, 0186 Oslo, Norway}
\affil[16]{Oslo Metropolitan University, Pilestredet 46, 0167 Oslo, Norway}
\affil[17]{Oslo University Hospital {Ullev{\aa}l}, Kirkeveien 166, 0450 Oslo, Norway}
\affil[18]{Machine \& Hybrid Intelligence Lab, Department of Radiology, Northwestern University, Chicago, USA}
\affil[*]{corresponding author: Sharib Ali (s.s.ali@leeds.ac.uk)}
\affil[$\dag$]{these authors contributed equally to this work}
 
\begin{abstract}
Polyps in the colon are widely known cancer precursors identified by colonoscopy. Whilst most polyps are benign, the polyp's number, size and surface structure are linked to the risk of colon cancer. Several methods have been developed to automate polyp detection and segmentation. However, the main issue is that they are not tested rigorously on a large multicentre purpose-built dataset, one reason being the lack of a comprehensive public dataset. As a result, the developed methods may not generalise to different population datasets. To this extent, we have curated a dataset from six unique centres incorporating more than 300 patients. The dataset includes both single frame and sequence data with 3762 annotated polyp labels with precise delineation of polyp boundaries verified by six senior gastroenterologists. To our knowledge, this is the most comprehensive detection and pixel-level segmentation dataset (referred to as \textit{PolypGen}) curated by a team of computational scientists and expert gastroenterologists. The paper provides insight into data construction and annotation strategies, quality assurance, and technical validation. Our dataset can be downloaded from \url{ https://doi.org/10.7303/syn26376615}. 
\end{abstract}
\begin{document}

\flushbottom
\maketitle
\thispagestyle{empty}
\section*{Background}
About 1.3 million new cases of colorectal cancer (CRC) are detected yearly in the world, with about 51\% mortality rate, and CRC is the third most common cause of cancer mortality~\cite{Bray2018}. Approximately, 90\% of CRCs result from slow transformation of the main benign precursors, adenomas or serrated polyps to CRC, but only a minority of them progress to CRC~\cite{leslie2002colorectal,loeve2004national}. It is particularly challenging to assess the malignant potential for lesions smaller than 10 mm. As a consequence most detected lesions are removed with subsequent CRC mortality reduction~\cite{kaminski2017increased}. The removal of the lesions depends also of an exact delineation of the boundaries to assure complete resection. If the lesions are detected and completely removed at a precancerous stage, the mortality is nearly null~\cite{Brenner2014crc}. Unfortunately, there is a considerable limitation related to various human skills ~\cite{hetzel2010variation,kahi2011prevalence} confirmed in a recent systematic review and meta-analysis demonstrating miss rates of 26\% for adenomas, 9\% for advanced adenomas and 27\% for serrated polyps~\cite{zhao2019magnitude}.
A thorough and detailed assessment of the neoplasia is essential to assess the  malignant potential and the appropriate treatment. This assessment is based on size, morphology and surface structure. Currently, the Paris classification, prone to substantial inter-observer variation even among experts, is used to assess the morphology~\cite{van2015polyp}. The surface structure classified by the Kudo pit pattern classification system or the Narrow-Band Imaging International Colorectal Endoscopic (NICE) classification system also help to predict the risk and degree of malignant transformation~\cite{saito2013multicentre}. This classification system may to some extent also predict the histopathological classification into either adenomas, sessile serrated lesions (SSLs), hyperplastic polyps or traditional serrated adenoma (TSA)~\cite{saito2013multicentre}. Unfortunately, these macroscopic classification systems are prone to substantial inter-observer variations, thus a high performing automatic computer-assisted system would be of great important both to increase detection rates and also reduce inter-observer variability. To develop such a system large segmented image databases are required. While current deep learning approaches has been instrumental in the development of computer-aided diagnosis (CAD) systems for polyp identification and segmentation, most of these trained networks suffer from huge performance gap when out-of-sample data have large domain shifts. On one hand, training models on large multi-centre datasets all together can lead to improved generalisation, but at an increased risk of false detection alarms~\cite{liu2020}. On the other hand, training and validation on centre-based splits can improve model generalisation. Most reported works are not focused on multi-centre data at all. This is mostly because of the lack of comprehensive multi-centre and multi-population datasets. In this paper, we present the \textit{PolypGen} dataset that incorporates colonoscopy data from 6 different centres for multiple patient and varied populations. Attentive splits are provided to test the generalisation capability of methods for improved clinical applicability. The dataset also is suitable for exploring federated learning and training of other time-series models. PolypGen can be pivotal in algorithm development and in providing more clinically applicable CAD detection and segmentation systems.

\begin{figure}[t!h!]
    \centering
    \includegraphics[width=0.9\textwidth]{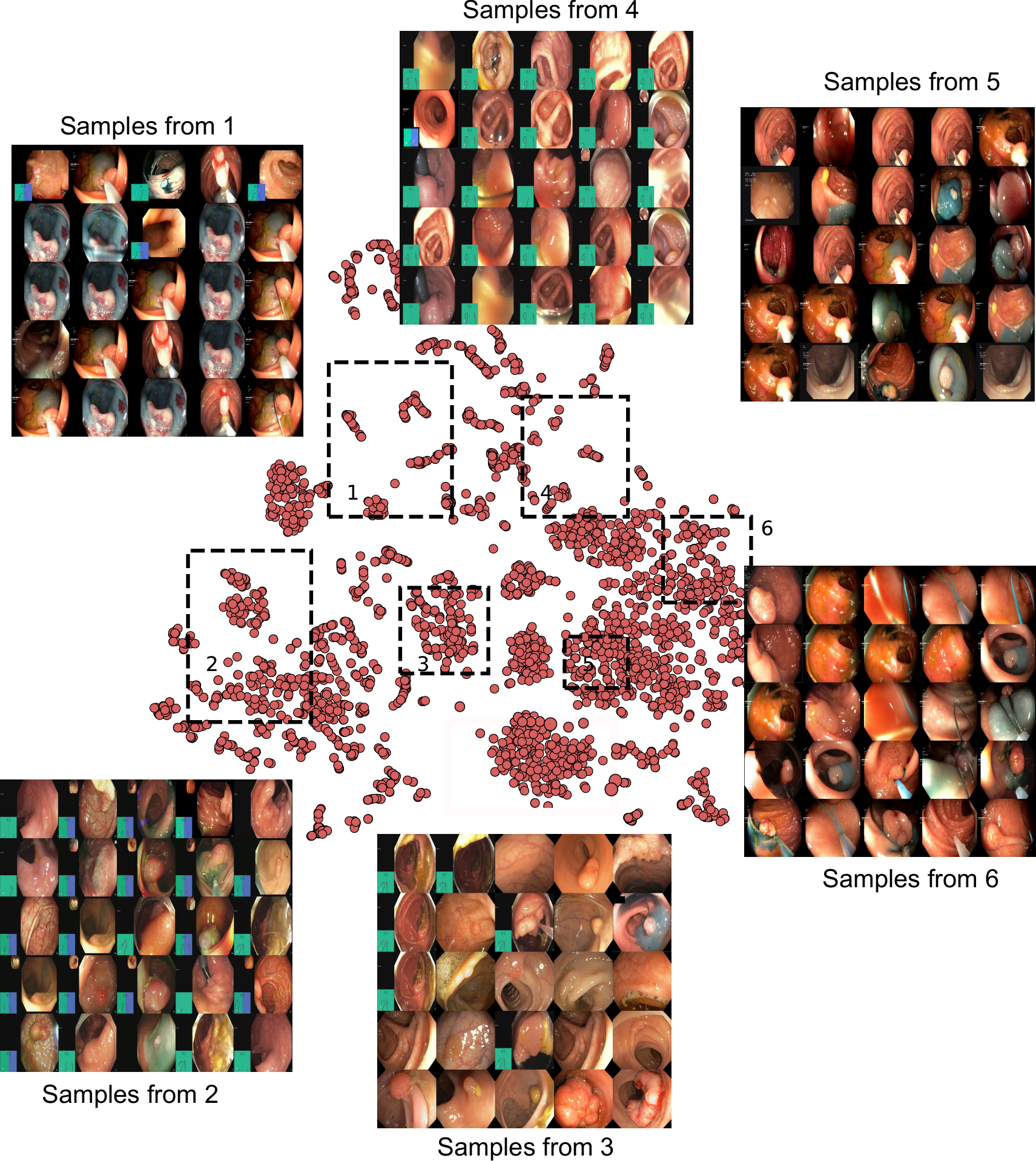}
    \caption{\textbf{t-SNE plot for positive samples:} 2D t-SNE embedding of the ``PolypGen'' dataset based on deep autoencoder extracted features. Each point is an image in the positive samples of the dataset. For each of the six boxed regions (dashed black lines) 25 images were randomly sampled for display in a $5\times 5$ image grid. Here, the 1st  boxed region represents mostly the sequence data. Interestingly, the 3rd, the 4th, and the 6th boxed regions mostly represent both polyp and non-polyp data and heterogeneously distributed. Samples from 2nd and the 5th boxed regions shows mostly protruded polyps but with differently positioned endoscopy locations. Some samples in these also include the colonoscopy frames with dyes.}
    \label{fig:supplementary_1}
\end{figure}

Although there are some publicly available datasets for colonoscopic single frames and videos (Table~\ref{tab:datasets}), lack of pixel-level annotations and preconditions applied for access of them pose challenges in its wide usability for method development. Many of these datasets are by request which requires approval from the data provider that usually takes prolonged time for approval and the approval is not guaranteed. Similarly, some datasets do not include pixel-level ground truth for the abnormality location which will cause difficulty in development or validation of CAD systems (\emph{e.g.}, El salvador atlas~\cite{elsalvadoratlas} and Atlas of GI Endoscope~\cite{atlas}). Moreover, most of the publicly available datasets include limited number of images frames from one or a few centres only (\emph{e.g.}, datasets provided in ~\cite{weo,mesejo2016computer,mesejo2016computer,ali2021_endoCV2020}). To this end, the presented \textit{PolypGen} dataset is composed of a total of {8037 frames} including both single and sequence frames. The provided comprehensive dataset consists of 3762 positive sample frames collected from six centres and {4275 negative sample frames} collected from four different hospitals. The \textit{PolypGen} dataset comprises of varied population data, endoscopic system and surveillance expert, and treatment procedures for polyp resections. A t-SNE plot for positive samples provided in the Fig.~\ref{fig:supplementary_1} demonstrates the diversity of the compiled dataset.
\begin{table*}[t!]
     \caption{\textbf{An overview of existing gastrointestinal (GI) lesion datasets including polyps:} number of images or videos along with the availability type is provided.} 
	\centering
	\begin{tabular}{l|l|l|l} 
		\hline
		\bf Dataset &\bf Findings & \bf Size & \bf Availability \\ \hline
		
		Kvasir-SEG~\cite{jha2020kvasir} & Polyps & 1000 images$^\dag$&  open academic\\ \hline
		
		HyperKvasir~\cite{borgli2020hyperkvasir} & GI findings including polyps & 110,079 images and 374 videos & open academic \\ \hline
		
		Kvasir-Capsule~\cite{smedsrud2021kvasir}& GI findings including polyps$^\diamond$ & 4,741,504 images & open academic \\ \hline
		
		CVC-ColonDB~\cite{bernal2012towards}& Polyps & 380 images$^\dag$ $^\dagger$&  by request$^\bullet$ \\ 
		\hline
		ETIS-Larib Polyp DB~\cite{silva2014toward} & Polyps & 196 images$^\dag$ &  open academic \\ \hline

	    EDD2020~\cite{ali2021_endoCV2020, EndoCV2020} & GI lesions including polyps  & 386 images & open academic\\ \hline 
		
		CVC-ClinicDB~\cite{bernal2015wm} & Polyps & 612 images$^\dag$&  open academic \\ \hline 
		
		CVC-VideoClinicDB~\cite{bernal2017miccai} & Polyps & 11,954 images$^\dag$&  by request$^\bullet$ \\ \hline

		ASU-Mayo polyp database~\cite{tajbakhsh2015automated} & Polyps & 18,781 images$^\dag$ & by request$^\bullet$  \\ \hline
		KID~\cite{koulaouzidis2017kid} & \begin{tabular}[c]{@{}l@{}}Angiectasia, bleeding,\\ inflammations$^\diamond$
 \end{tabular} & \begin{tabular}[c]{@{}l@{}}2371 images,\\ 47 videos \end{tabular}  & open academic$^\bullet$\\ \hline
	   Atlas of GI Endoscope~\cite{{atlas}}& GI lesions & 1295 images & unknown$^\bullet$ \\   \hline
	   El salvador atlas~\cite{elsalvadoratlas} & GI lesions &  5071 video clips & open academic$^\clubsuit$ \\ \hline
	   \textbf{PolypGen (Ours)}~\cite{EndoCV2021,synapsePolypGen} & Multi-centre colon polyps & 1537 images$^\dag$\& 2225 video sequence & open academic \\ \hline
	   \multicolumn{4}{l}{$^\dag$Including ground truth segmentation masks \hspace{.1cm}
	   $^\ddagger$Contour \hspace{.1cm}
	   $^\diamond$Video capsule endoscopy\hspace{.1cm}
	   $^\bullet$Not available anymore}\\
	    \multicolumn{4}{l}{$^\clubsuit$Medical atlas for education with several low-quality samples of various GI findings.
	    \hspace{.1cm}}\\
	 \end{tabular}
	\label{tab:datasets}
\end{table*}
\begin{figure}[t!]
    \centering
    \includegraphics[width=0.90\textwidth]{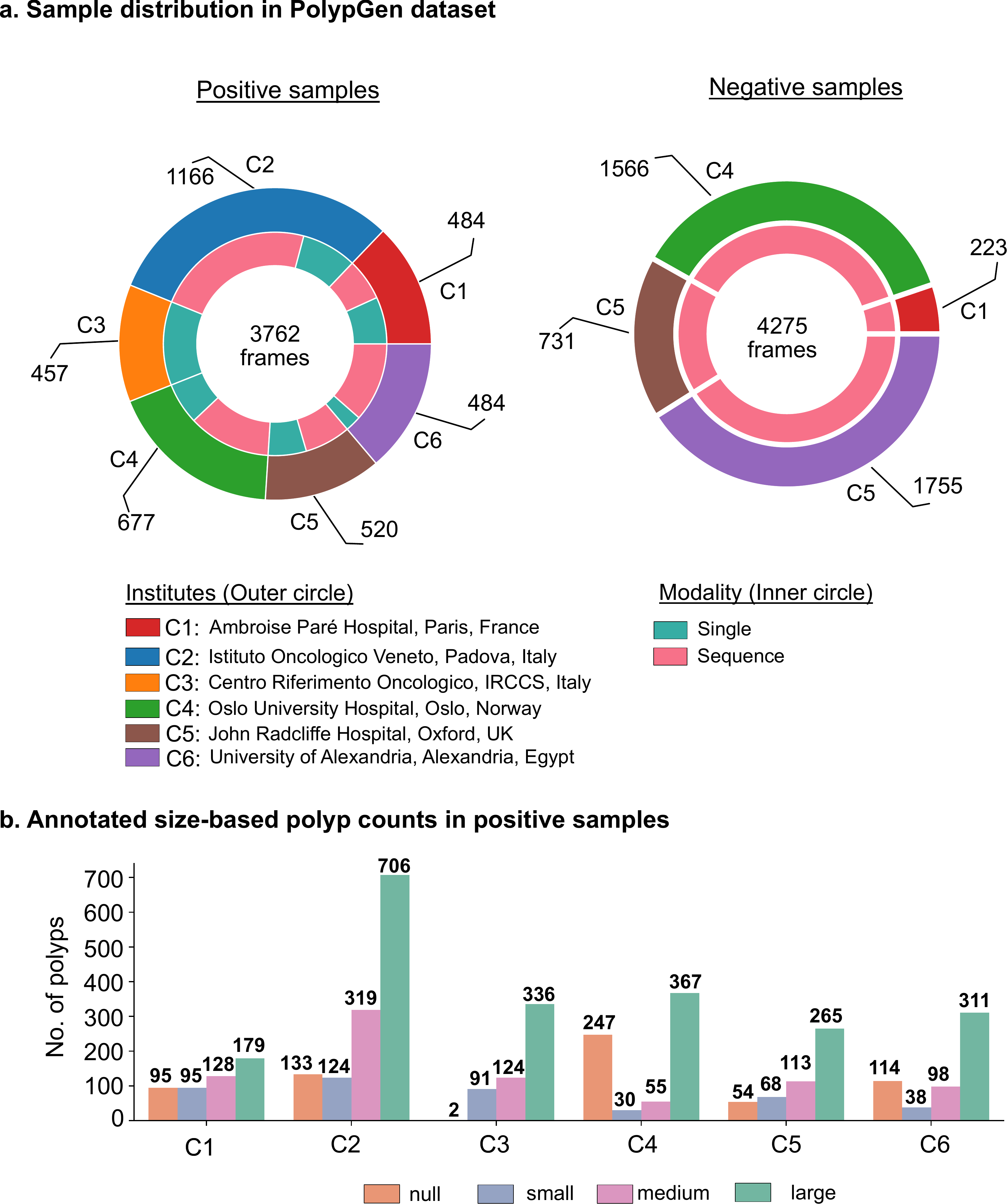}
    \caption{\textbf{PolypGen dataset:} (a) Positive (both single and sequence frames) and negative samples (sequence only) from each centre, and (b) polyp size-based histogram plot for positive samples showing variable sized annotated polyps in the dataset (small is $\leq 100 \times 100$ pixels; medium is $> 100 \times 100 \leq 200 \times 200$, and large is $> 200 \times 200$ pixels). Null represents no polyp present in the sample.}
    \label{fig:polypGenDatasetOverview}
\end{figure}
\section*{Methods}
\subsection*{Ethical and privacy aspects of the data}
\begin{table*}[t!]
\caption{\textbf{Data collection information for each centre:} Data acquisition system and patient consenting information. \label{tab:patientConsentingInfo}}
\small{
\begin{tabular}{l|l|l|l}
\hline
\textbf{centres}                              & \textbf{System info.}                & \textbf{Ethical approval} & \textbf{Patient consenting type} \\ \hline
Ambroise Par\'{e} Hospital, Paris, France         & Olympus Exera 195              & N° IDRCB:   & Endospectral study   \\ 
& & 2019-A01602-55&\\ \hline
Istituto Oncologico Veneto, Padova, Italy     & Olympus endoscope H190               & NA                        & Generic patients consent         \\ \hline
Centro Riferimento Oncologico, IRCCS, Italy   & Olympus VG-165, CV180, H185          & NA                        & Generic patients consent         \\ \hline
Oslo University Hospital, Oslo, Norway & \begin{tabular}[c]{@{}l@{}}Olympus Evis Exera III, CF 190 \end{tabular} &Exempted$^\dag$  & Written informed consent      \\ \hline
John Radcliffe Hospital, Oxford, UK           & GIF-H260Z, EVIS Lucera CV260,  & REC Ref:       & Universal consent \\
&Olympus Medical Systems & 16/YH/0247 &
\\ \hline
University of Alexandria, Alexandria, Egypt & Olympus Exera 160AL, 180AL                             & NA                        & Written informed consent         \\ \hline

\multicolumn{4}{l}{$^\dag$ Approved by the data inspectorate. No further ethical approval was required as it did not interfere with patient treatment} \\
\end{tabular}
}

\end{table*}
Our multi-centre polyp detection and segmentation dataset consists of colonoscopy video frames that represent varied patient population imaged at six different centres including Egypt, France, Italy, Norway and the United Kingdom (UK). Each centre was responsible for handling the ethical, legal and privacy of the relevant data. The data collection from each centre included either two or all essential steps described below:
\begin{itemize}
\item[1.] Patient consenting procedure at each individual institution (required)
\item[2.] Review of the data collection plan by a local medical ethics committee or an institutional review board
\item[3.] Anonymization of the video or image frames (including demographic information) prior to sending to the organizers (required)
\end{itemize}
\noindent Table~\ref{tab:patientConsentingInfo} illustrates the ethical and legal processes fulfilled by each centre along with the endoscopy equipment and recorders used for the data collected.
\subsection*{Study design}
PolypGen data was collected from 6 different centres. More than 300 unique patient videos/frames were used for this study. The general purpose of this diverse dataset is to allow robust design of deep learning models and their validation to assess their generalizability capability. In this context, we have proposed different dataset configurations for training and out-of-sample validation and proposed unique generalization assessment metrics to reveal the strength of deep learning methods. Below we provide a comprehensive description of dataset collection, annotation strategies and its quality, ethical guidelines and metric evaluation strategies. 
\subsection*{Video acquisition, collection and dataset construction}\label{sec:data_collection}
A consortium of six different medical data centres (hospitals) were built where each data centre provided videos and image frames from at least 50 unique patients. The videos and image samples were collected and sent by the senior gastroenterologists involved in this project. The collected dataset consisted of both polyp and normal mucosa colonoscopy acquisitions. To incorporate the nature of polyp occurrences and maintain heterogeneity in the data distribution, the following protocol was adhered for establishing the dataset:
\begin{itemize}
    \item[1.] Single frame sampling from each patient video incorporated different view points
    \item[2.] Sequence frame sampling consisted of both visible and invisible polyp frames (at most cases) with a minimal gap
    \item[3.] While single frame data consisted of all polyp instances in that patient, sequence frame data consisted of only a localised targeted polyp 
    \item[4.] Positive sequence included both positive and negative polyp instances but from video with confirmed polyp location while for negative sequence only patient videos with normal mucosa were used
\end{itemize}
An overview of the number of samples comprising positive samples and negative samples is presented in Fig.~\ref{fig:polypGenDatasetOverview} a. The total positive samples of 3762 frames are released that comprises of 484, 1166, 457, 677, 458 and 520 frames from centres C1, C2, C3, C4, C5 and C6, respectively. These frames consist of 1537 single frames (1449 frames from C1-C5 also provided in EndoCV2021 challenge and 88 frames from C6),  and 2225 sequence frames with majority of sequence data sampled from centres C2 (865), C4 (450), and C6 (432). The number of polyp counts for pixel-level annotation of small ($\leq 100 \times 100$), medium (between $> 100 \times 100$ pixels and $ \leq 200 \times 200$ pixels), large ($\geq 200 \times 200$ pixels) sized polyps from each centre including no polyp frames but frames in close proximity of polyp are represented as histogram plot (Fig.~\ref{fig:polypGenDatasetOverview} b). The total annotations for polyp that we release is 3447. All these polyp samples are verified by expert gastroenterologists. 

We have provided both still image frames and continuous short video sequence data with their corresponding annotations. The positive and negative samples in the dataset of the polyp generalisation (PolypGen) are further detailed below.
\begin{figure}[t!h!]
    \centering
    \includegraphics[width=0.6\textwidth]{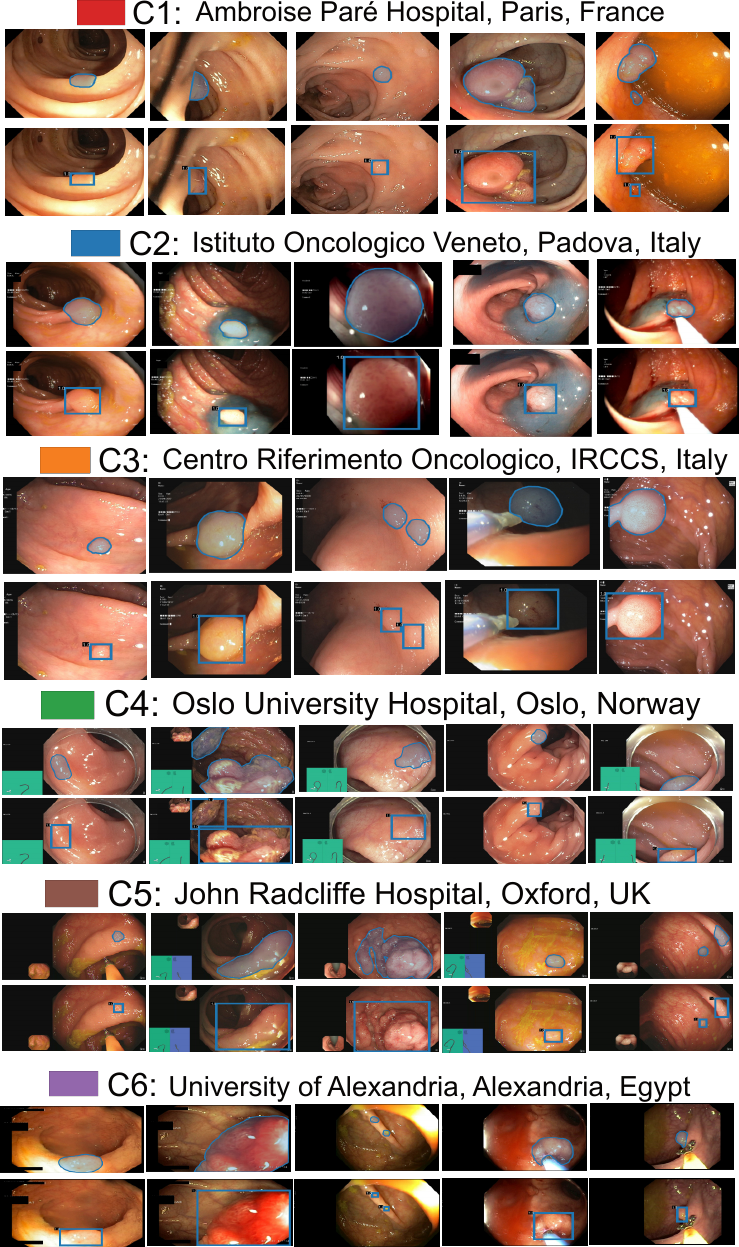}
    \caption{\textbf{Sample polyp annotations from each centre:} Segmentation area with boundaries and corresponding bounding box/boxes overlaid images from all six centres. Samples include both small sized polyp ($<10000~\text{pixels}$) including some flat polyp samples to large sized ($\geq~\text{40000 pixels}$) polyps and polyps during resection procedure such as polyps with blue dyes.}
    \label{fig:supplementary_2}
\end{figure}
\subsubsection*{Positive samples}
\begin{figure}[t!]
    \centering
    \includegraphics[width=0.9\textwidth]{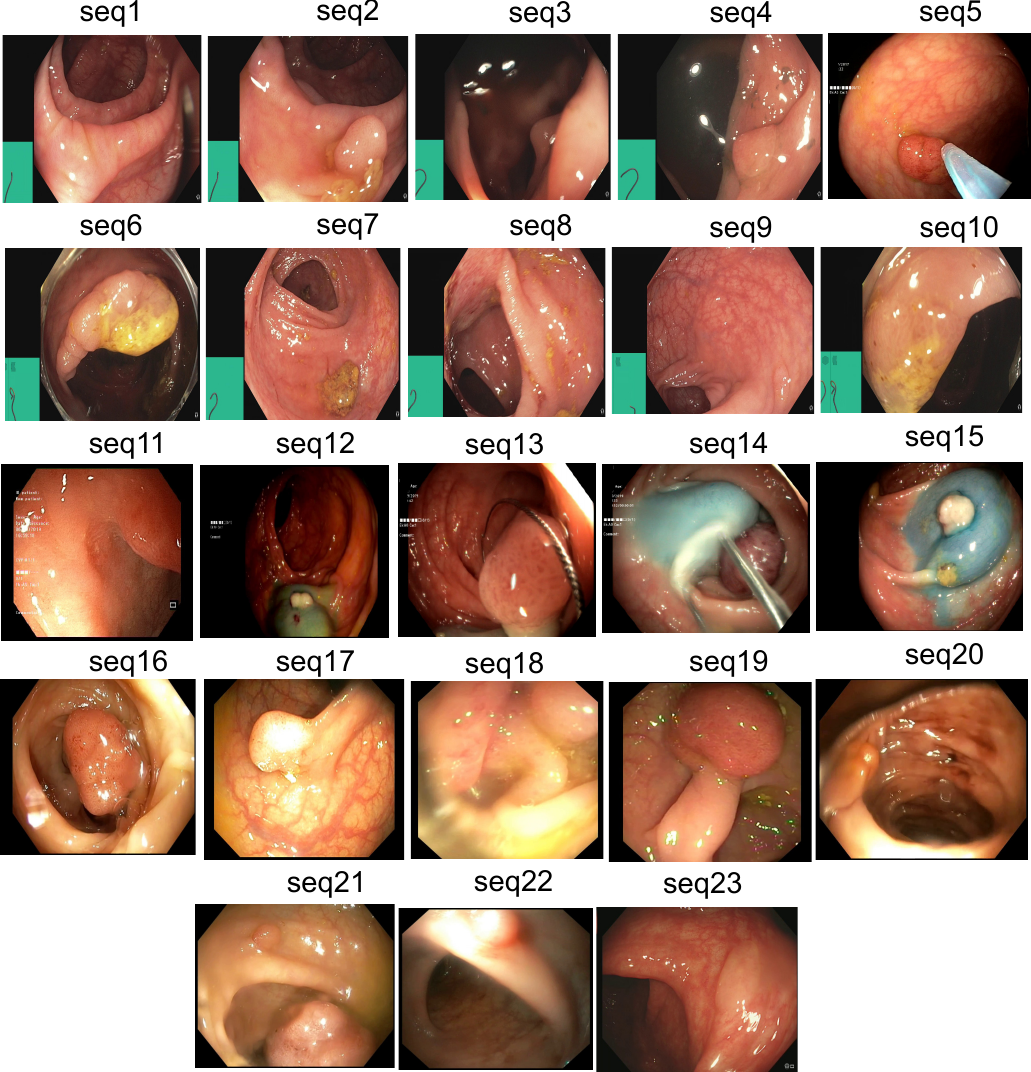}
        \caption{\textbf{Positive sequence data:} {Representative samples chosen from 23 sequences of the provided positive samples data. A summary description is provided in Table~\ref{tab:positiveSamples}. Parts of images have been cropped for visualization. }\label{fig:positive_samples}}
\end{figure}

Positive samples consist of video frames from the patient with a diagnosed polyp case. The selected frames may or may not have the polyp in it but may be located near to the chosen frame. Nevertheless, a majority of these frames consists of at least one polyp in the frame. For the sequence positive samples, the continuity of the appearance and disappearance of the polyp similar to real scenario has been taken into account and thus these frames can have a mixture of polyp instances and frames with normal mucosa. Table \ref{tab:positiveSamples} is provided to detail the characteristics of 23 sequence data included in our dataset. It can be observed from Figure~\ref{fig:positive_samples} that varied sized polyps are included in the dataset with variable view points, occlusions and instruments. Exemplary pixel-level annotations of positive polyp samples for each centre and their corresponding bounding boxes are presented in Fig.~\ref{fig:supplementary_2}. 
\begin{table}[t!]
\caption{\textbf{Positive sample sequence summarised attribute:} {Total of 23 sequences are provided as positive sample sequence for patients with polyp instances during colonoscopy examination. Here JNET refers to Japan NBI Expert Team classification score. These sequences depict different sized polyps and location with different artifacts and varying visibility. Sequences referring to one selected image is shown in Fig.~\ref{fig:positive_samples}\label{tab:positiveSamples}.}}
\centering
\begin{tabular}{l|l|l}
\textbf{Sequence}  & \textbf{Description}                                                & \textbf{Artifact}                                                           \\ \hline
seq1  & Normal mucosa                                              & Light reflections; green patch\\
seq2  & 5 mm polyp at 6 o’clock                                            & Partially covered with stool; reflections; green patch                                   \\
seq3  & Polyp at distance, 4 o’clock                                & Light reflection from liquid; green patch                                                                                     \\
seq4  & 2-3 mm polyp                                             & Liquid covering half of the image; green patch                                                                        \\
seq5  & 5 mm polyp catched by a snare                                           & Partial occlusion by biopsy instrument                                                                                   \\
seq6 & Polyp covering half of the circumference                                          & Cap; green patch                                     \\
seq7  & Normal mucosa                                           & Light reflection; some remnant stool; green patch                                                                            \\
seq8  & Typical flat cancer                                             & Light reflection; green patch                                       \\
seq9  & 2 mm polyp at 2 o’clock                                           & Light reflection; green patch                                       \\
seq10 & Subtle small protrusions   &  some remnant stool                                                                  \\
seq11 & Polyp at 2-3 o’clock                                           & Light reflections in the periphery              
\\
seq12 & Dye lifted 4-5 mm polyp &   Low contrast        \\
seq13 & 6-7 mm polyp catched with a snare   & Low contrast; small reflections\\
{
seq14} & {Paris 1 p polyp, large long stak, JNET 2a} & {Lifted by Indigo Carmine, snare place around the stalk}\\
{seq15} & {Paris 1 s JNET2a polyp and 1 Paris 1 sp to the left} & {Lifted by Indigo Carmine}\\
{seq16} & {Paris 1 p polyp, large long stalk, JNET 2a} & {Lifted by Indigo Carmine}\\
{seq17} & {Paris 1 sp polyp} & {Light reflections make surface assessment impossible}\\
{seq18} & {Difficult interpretation} & {Blurry image and reduced view}\\
{seq19} & {Paris 1 p polyp, large long stalk, JNET 2a} & {Less contrast and slightly occluded}\\
{seq20} & {half of the polyp visible} & {Blurry image, with some blood on the mucosa}\\
{seq21} & {Two adenomas polyp} & {Blurry image}\\
{seq22} & {adenomas polyp}  & {Blurry image makes exact diagnosis impossible}\\
{seq23} & {serrated polyp}  & {Perfect clean mucosa, minor light reflections}\\
\hline
\end{tabular}
\end{table}
%
\subsubsection*{Negative samples}
\begin{figure}[t!]
    \centering
    \includegraphics[width=0.9\textwidth]{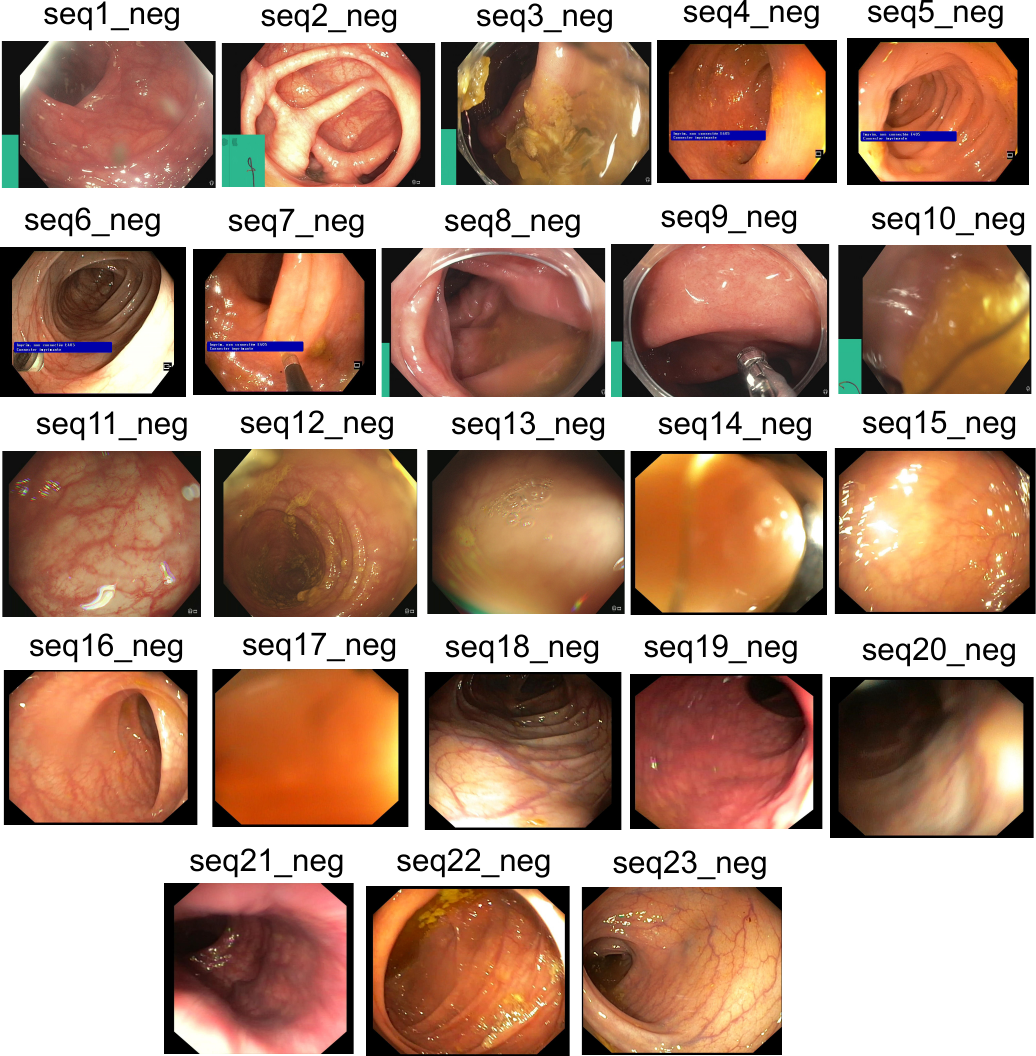}
        \caption{\textbf{Negative sequence data:} {Representative samples chosen from each sequence of the provided negative samples data. A summary description is provided in Table~\ref{tab:negativeSamples}. Parts of images have been cropped for visualization.}  \label{fig:negative_samples}}
\end{figure}

Negative samples mostly refer to the negative sequences released in this dataset, i.e. no polyp frames. These sequences are taken from patient videos which consisted of confirmed absence of polyps (\emph{i.e.}, normal mucosa) in the acquired videos {or at areas away from the polyp occurrences.} It includes cases with anatomies such as colon linings, light reflections and mucosa covered with stool that may be confused with polyps (see Figure~\ref{fig:negative_samples} and corresponding negative sequence attributes in Table~\ref{tab:negativeSamples}).
\begin{table}[t!]
\caption{\textbf{Negative sample sequence summarised attribute:} Total of 23 sequences are provided as negative sample sequence for patients with no polyp during colonoscopy examination. These sequences depict different artifacts and varying visibility of vascular pattern and occlusion of mucosa.\label{tab:negativeSamples}}
\centering
\begin{tabular}{l|l|l}
\textbf{Sequence}  & \textbf{Description}                                                & \textbf{Artifact}                                                           \\ \hline
seq1\_neg  & Normal vascular pattern                                             & Light reflections in the periphery; not clean lens                   \\
seq2\_neg  & Normal vascular pattern                                             & Contracted bowel; light reflections                                   \\
seq3\_{neg}  & Mucosa not satisfactory visulaized                                  & Stool covers the field of view                                                                                    \\
seq4\_neg  & Reduced vascular pattern                                            & Light reflections and small amount of stool                                                                       \\
seq5\_neg  & Reduced vascular pattern                                            & Light reflections                                                                                                 \\
seq6\_neg  & Normal vascular pattern                                             & Light reflections; biopsy forceps                                     \\
seq7\_neg  & Normal vascular pattern                                             & Very close to the luminal wall                                                                                    \\
seq8\_neg  & Normal vascular pattern                                             & Blurry; semi-opaque liquid; cap                                       \\
seq9\_neg  & Normal vascular pattern                                             & Blurry; semi-opaque liquid; cap                                         \\
seq10\_neg & Not possible to assess the mucosa                       & Blurry; occluded                                                                                                  \\
seq11\_neg & Normal vascular pattern                                             & Light reflections in the periphery; bubble on the lens               \\
seq12\_neg & Normal vascular pattern                                             & Not clean lens, mucosa covered by stool                                                                           \\
seq13\_neg & \begin{tabular}[c]{@{}l@{}}Probably normal vascular pattern;\\  Not possible to assess the mucosa\end{tabular}    & \begin{tabular}[c]{@{}l@{}}Air bubbles; remnant stool; \\ too close to the mucosa, blur, reflections\end{tabular}\\

{seq14\_neg} & {\begin{tabular}[c]{@{}l@{}}Clean bowel, normal vascular pattern\end{tabular}}   & {\begin{tabular}[c]{@{}l@{}}Very close to the mucosa in all\end{tabular}}\\
{seq15\_neg} & {\begin{tabular}[c]{@{}l@{}}Clean bowel, normal vascular pattern\end{tabular}}    & {\begin{tabular}[c]{@{}l@{}}Some bubbles and light reflections\end{tabular}}\\
{seq16\_neg} & {\begin{tabular}[c]{@{}l@{}}Clean bowel, normal vascular pattern\end{tabular}}   & {\begin{tabular}[c]{@{}l@{}}Some bubbles and light reflection\end{tabular}}\\
{seq17\_neg} & {\begin{tabular}[c]{@{}l@{}}Clean bowel, normal vascular pattern\end{tabular}}    & {\begin{tabular}[c]{@{}l@{}}Very close to the mucosa in all\end{tabular}}\\
{seq18\_neg} & {\begin{tabular}[c]{@{}l@{}}Clean bowel, normal vascular pattern, well distended\end{tabular} }   & {\begin{tabular}[c]{@{}l@{}}Some stool residues\end{tabular}}\\
{seq19\_neg} &{\begin{tabular}[c]{@{}l@{}}Clean bowel, normal vascular pattern, well distended\end{tabular}}   & {\begin{tabular}[c]{@{}l@{}}Some liquid residues\end{tabular}}\\
{seq20\_neg} & {\begin{tabular}[c]{@{}l@{}}Clean bowel, normal vascular pattern, well distended\end{tabular}}    & {\begin{tabular}[c]{@{}l@{}}Some stool residues and reflections\end{tabular}}\\
{seq21\_neg} & {\begin{tabular}[c]{@{}l@{}}Clean bowel, normal vascular pattern\end{tabular}}   & {\begin{tabular}[c]{@{}l@{}}Very close, minor stool residues in last images\end{tabular}}\\
{seq22\_neg} & {\begin{tabular}[c]{@{}l@{}}Clean bowel, normal vascular pattern, well distended\end{tabular}}    & {\begin{tabular}[c]{@{}l@{}}Some liquid and stool residues, reflections\end{tabular}}\\
{seq23\_neg} & {\begin{tabular}[c]{@{}l@{}}Perfect clean bowel, normal vascular pattern, well distended\end{tabular}}   & {\begin{tabular}[c]{@{}l@{}}Some light reflections\end{tabular}}\\
\hline
\end{tabular}
\end{table}                                                    
\subsection*{Annotation strategies and quality assurance}
\begin{figure}[t!]
    \centering
    \includegraphics[scale=0.45]{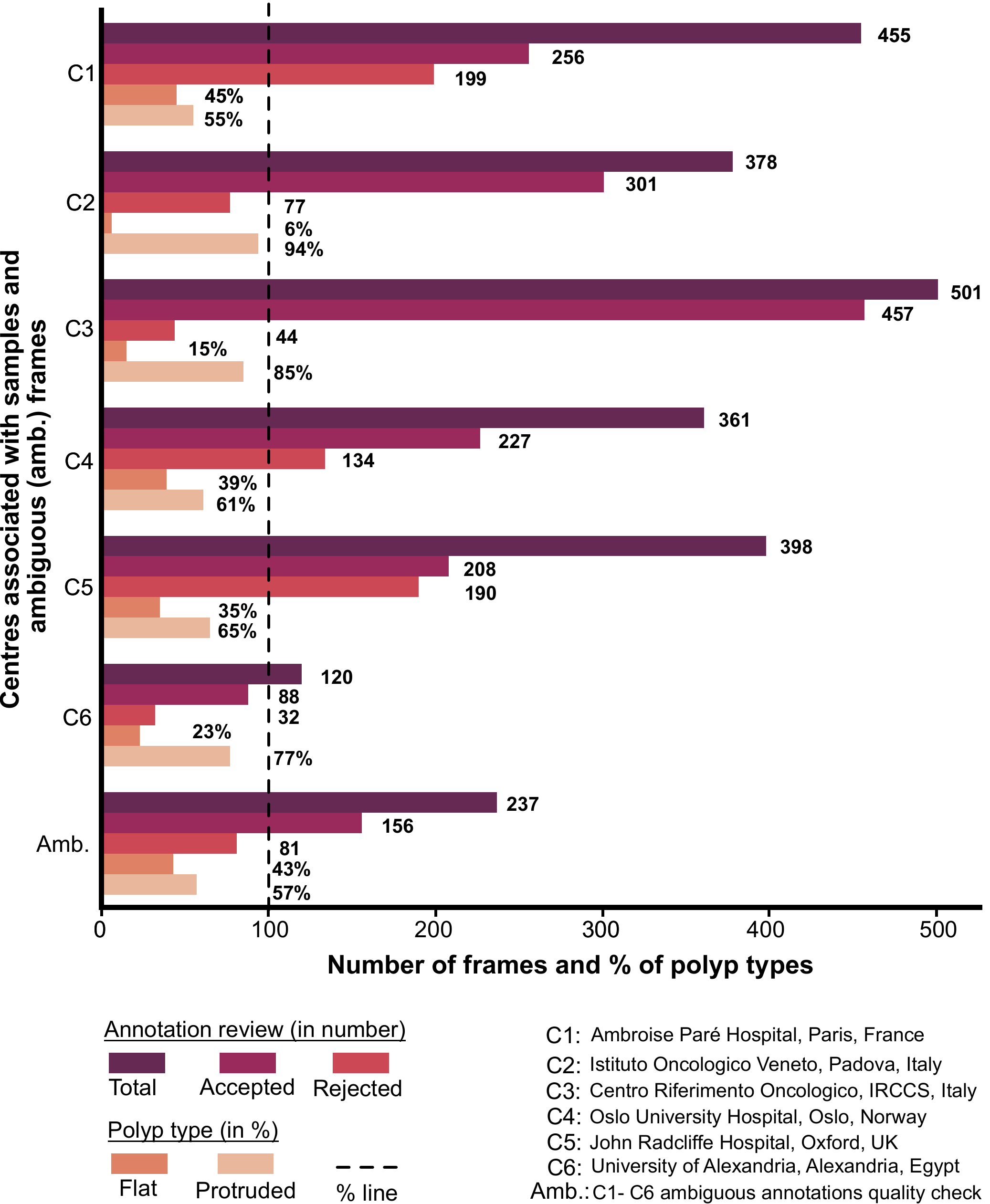}
    \caption{\textbf{Annotation quality review:} Total curated frames along with accepted and rejected frame numbers during annotation quality review by experts for single frame data. Annotated frames with \% of flat and protruded polyps categorised during annotation are also provided.}
    \label{fig:qualityAssurance}
\end{figure}

A team of 6 senior gastroenterologists (all over 20 years of experience in endoscopy), two experienced post-doctoral researchers, and one PhD student were involved in the data collection, data sorting, annotation and the review process of the quality of annotations. For details on data collection and data sorting please refer to Section~\textbf{Video acquisition, collection and dataset construction}. All annotations were performed by a team of three experienced researchers using an online annotation tool called Labelbox\footnote{https://labelbox.com}. {The dataset was divided equally between the three reviewers for the annotation process where each research annotated a specific group of frames. However, all the annotated frames were revised by all the senior gastroenterologists team}. 
%
Each annotation was later cross validated for accurate segmentation margins by the team and by the centre expert. Further, an independent binary review process was then assigned to a senior gastroenterologists, in most cases experts from different centres were assigned.
%
A protocol for manual annotation of polyp was designed to minimise the heterogeneity in the manual delineation process. 
{The protocol was in detail discussed together with the clinical experts and the annotators during several weekly meeting. Here, we only present a brief on the important aspects of the annotation that should be taken care during annotations. Example samples were provided by expert endoscopists to the annotators especially this was the case in the video annotations.} The set protocols are listed below (refer~Fig.~\ref{fig:supplementary_2} for final ground truth annotations):
\begin{itemize}
    \item Clear raised polyps: Boundary pixels should include only protruded regions. Precaution has to be taken when delineating along the normal colon folds 
    \item Inked polyp regions: Only part of the non-inked appearing object delineation
    \item Polyps with instrument parts: Annotation should not include instrument and is required to be carefully delineated and may form more than one object
    \item Pedunculated polyps: Annotation should include all raised regions unless appearing on the fold
    \item Flat polyps: Zooming the regions identified with flat polyps before manual delineation. Also, consulting centre expert if needed.
    {\item Video sequence annotation: One sample from expert gastroenterologist were provided for sequences that showed difficulty in distinguishing between mucosa and polyp. Polyps that are distant and not clearly visible were also not annotated as polyps.
    \item Tackling with occlusion: Polyps that were occluded with stool or instrument were required to exclude the parts of mucosa that were obstructed.
    \item Cancerous mucosa: Mucosa that were already cancerous but not appear as polyps were excluded from the annotation. However, raised mucosal surface that charactised adenomatous polyps were included.}
\end{itemize}
Each of these annotated masks were reviewed by expert gastroenterologists. During this review process, a binary score was provided by the experts depending on whether the annotations were clinically acceptable or not. Some of the experts also provided feedback on the annotation and these images were placed into ambiguous category for further rectification based on expert feedback. These ambiguous category was then jointly annotated by two researchers and further sent for review to one expert. The outcome of these quality checks are provided in Figure~\ref{fig:qualityAssurance}. It can be observed that large fraction (30.5\%) of annotations were rejected (excluding ambiguous batch, total annotations were 2213, among which only 1537 were accepted and 676 frames were rejected). Similarly, the ambiguous batch that included correction of annotations after the first review also recorded 34.17\% rejected frames on the second review.   
\section*{Data Records}
\begin{figure}[t!]
    \centering
    \includegraphics[width=1.0\textwidth]{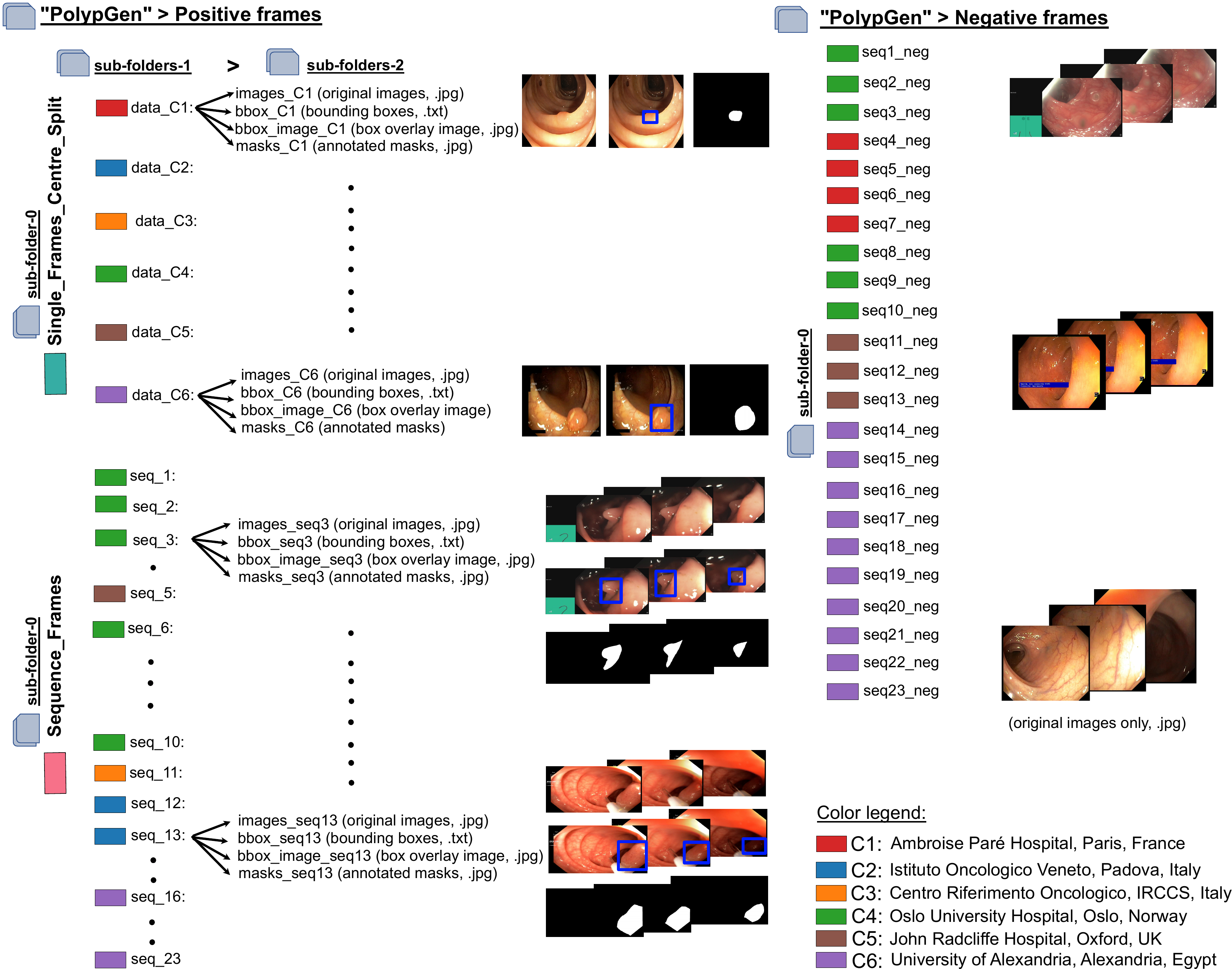}
    \caption{\textbf{Folder structure:} PolypGen dataset is divided into two folders -- positive frames and negative frames. Later, it is divided into three different levels for positive frames and only one level for negative samples. (On left) Sub-folder structure with different folder names and the format of data present in each sub-folder 2 is provided. Sample images are also shown. (On right) Sub-folders present in negative folder is shown with two sample sequences (from centre 4 and centre 1). Each data source centre is shown as color legends.}
    \label{fig:datareecords_folderStructure}
\end{figure}
A sub-set of this dataset ( from C1 - C5 except C6) forms the dataset of our EndoCV2021 challenge~\cite{Ali2022AssessingGO} (\textit{Addressing generalisability in polyp detection and segmentation}) training data (\url{https://endocv2021.grand-challenge.org}), \emph{i.e.}, an event held in conjunction with the IEEE International Symposium on Biomedical Imaging (ISBI 2021), Nice, France. The current released data consists of additional positive and negative frames for both single and sequence data and a 6th centre data (C6). The presented version does not consists of training and test splits and users are free to apply their own strategies as applicable to the nature of their work. To access the complete dataset, users are requested to create a Synapse account (\url{https://www.synapse.org/}) and then the compiled dataset can be downloaded at (\url{https://www.synapse.org/#!Synapse:syn45200214})\cite{synapsePolypGen} which  has been published under Creative Commons 4.0 International (CC BY) licence. Dataset can only be used for educational and research purposed and must cite this paper. All collected data has been obtained through a written patient consent or through an ethical approval as tabulated in Table~\ref{tab:patientConsentingInfo}.

The folder structure of the compiled multi-centre polyp detection and segmentation dataset is as presented in Figure~\ref{fig:datareecords_folderStructure}. The main folder ``PolypGen'' is divided into folders a) Positive and b) Negative. The positive folder is then subdivided (sub-folder level 0) into ``Single\_Frames\_centre\_Split'' and ``Sequence\_Frames''. Each of these folders are then further subdivided (sub-folder-1). Single frame data is further categorized as centre-wise split ``data-C1'' to ``data-C6'', where C1 representing centre 1 and C6 representing the 6th centre, while sequence frames are categorized into ``seq\_1'' to ``seq\_23'' (legend color in Figure~\ref{fig:datareecords_folderStructure} represents their corresponding centre). Finally, a second sub-folder level 2 includes four folders consisting of original images (.jpg format), annotated masks (.jpg format), bounding boxes (.txt) in the standard PASCAL VOC format~\cite{PASCALVOC:2012}, and images with box overlay (.jpg). No mix of centrewise sequence is done in any of the sequence data, and the data can consist of both positive negative polyp image samples. No polyp in positive samples mask empty bounding box files and mask with null values. Both masks and bounding box overlaid images are of same size as that of the original images. The negative frames folder on the other hand does consist of only sequence frames, i.e., sub-folder level 0 only as these sequence samples consists of patients with no polyps during the surveillance. To further assist the users, we have also provided the folder structure inside the main folder ``PolypGen''. The size of images provided in the dataset can range from $384\times 288$ pixels to  $1920 \times 1080$ pixels. The size of masks correspond to the size of the original images, however, the polyp sizes in the provided dataset is variable (as indicated in Figure~\ref{fig:polypGenDatasetOverview}). {{Since we followed a full anonymisation protocol, no gender or age information is provided.}}
\section*{Technical Validation}
For the technical validation, we have included single frames data (1449 frames) from five centres (C1 to C5) in our training set and tested on out-of-sample C6 data on both single (88 frames) and sequence frames (432 frames). Such out-of-sample testing on a completely different population and endoscopy device data allows to comprehensively provide evidence of generalisability of current deep learning methods. The training set was split into 80\% training only and 20\% validation data. Here, we take most commonly used methods for segmentation in the biomedical imaging community~\cite{long2015fully,ronneberger2015u,zhao2017pyramid,chen2018encoder,he2016deep}, including that for polyps. For reproducibility, we have included the train-validation split as \textit{.txt} files as well in the \textit{PolypGen} dataset folder. However, users can choose any set of different combined training or split training schemes for generalisation tests as they prefer, e.g., training on random three centres and testing on remaining three centres. Also, the dataset is suitable for federated learning (FL) approaches~\cite{Sashank21} that uses decentralised training and allow to aggregate the weights from the central server for improved and generalisable model without compromising data privacy.
\subsection*{Benchmarking of state-of-the-art methods}
To provide generalisation capability of some state-of-the-art (SOTA) methods, we have used a set of popular and well-established semantic segmentation CNN models on our PolypGen dataset. Each model was run for nearly 500 epochs with batch size of 16 for image size of $512\times512$. All models were optimised using Adam with weight decay of 0.00001 learning rate of 0.01 and allowing the best model to be saved after 100 epochs. Classical augmentation strategies were used that included scaling (0.5, 2.0), random cropping, random horizontal flip and image normalisation. All models were run on Quadro RTX6000. 
\subsubsection*{Evaluation metrics for segmentation.} We compute standard metrics used for assessing segmentation performances that includes Jaccard Index ($\text{JI} = \frac{TP} {TP + FP + FN}$), F1-score (aka Dice similarity coefficient, DSC), F2-score, precision (aka positive predictive value, PPV or $p={\frac {TP}{TP+FP}}$), recall ($r={\frac {TP}{TP+FN}}$), and overall accuracy ($\text{Acc.} ={\frac{TP + TN} {TP + TN + FP + FN}}$) that are based on  true positives (TP), false positives (FP), true negatives (TN), and false negatives (FN) pixel counts. Precision-recall tradeoff is also given by the Dice similarity coefficient (DSC) or F1-score and F2-scores:
\begin{equation}{\label{eq:fscore}}
\centering
    \mathrm{F_\beta} = (1 + \beta^2) \cdot \frac{{p} \cdot {r}}{(\beta^2 \cdot {p}) + {r}}, 
\end{equation}
where $\mathrm{F_\beta}$-score is computed as weighted harmonic means of precision and recall. 

Another commonly used segmentation metric that is based on the distance between two point sets, here ground truth (G) and estimated or predicted (E) pixels, to estimate ranking errors is the average Hausdorff distance ($d_{AHD}$) and defined as:
\begin{equation}
    d_{AHD}(G, E) = \bigg(\frac{1}{|G|} \sum_{g\in G} \min_{e\in E} d(g, e) + \frac{1}{|E|} \sum_{e\in E} \min_{g\in G} d (g, e)\bigg)/2.
\end{equation}

{Since boundary-distance-based metrics are insensitive to
the object size and sensitive to the object shape, we include two additional metrics which are average surface distance (ASD) and  Normalised surface dice (NSD). ASD is the average of all distances (Euclidean) between pixels on the
predicted object segmentation border and its nearest neighbour on the reference segmentation border. All obtained distances are averaged, yielding an average distance
value ASD for symmetric case is:}
\begin{equation}
    ASD{(G, E)} = \frac{1}{|G|+|E|} \bigg(\sum_{g\in G} \min_{e\in E} d(g, e) + \sum_{e\in E} \min_{g\in G} d (g, e)\bigg).
\end{equation}

{The normalized surface dice (NSD)~\cite{Nikolov2021} computes the fractional correctly predicted segmentation boundary using an additional threshold accounting for amount of class-specific distance deviation. In our experiments we have set it to 10. If $d$ be the distance and $\tau$ be the acceptable deviation (threshold) with $d({B_E,B_G})$ be the computed distance for predicted mask boundary w.r.t the nearest-neighbour distances to the reference segmentation boundary then for symmetric case NSD is given by:}
\begin{equation}
    d^{'}({B_E, B_G}) = \{d \in d(B_E, B_G) ~|~ d\leq \tau\} \quad \text{and} \quad NSD = \frac{|d^{'}({B_E, B_G})|+|d^{'}({B_G, B_E})|}{|d(B_E, B_G)| + |d(B_G, B_E)|}.
\end{equation}
\noindent {{NSD is bounded between 0 and 1 where 1 refers to the segmentation boundary below deviation threshold $\tau$.}}
%
%
\subsubsection*{Polyp segmentation benchmarking}
\begin{figure}[t!]
    \centering
    \includegraphics[width=1\textwidth]{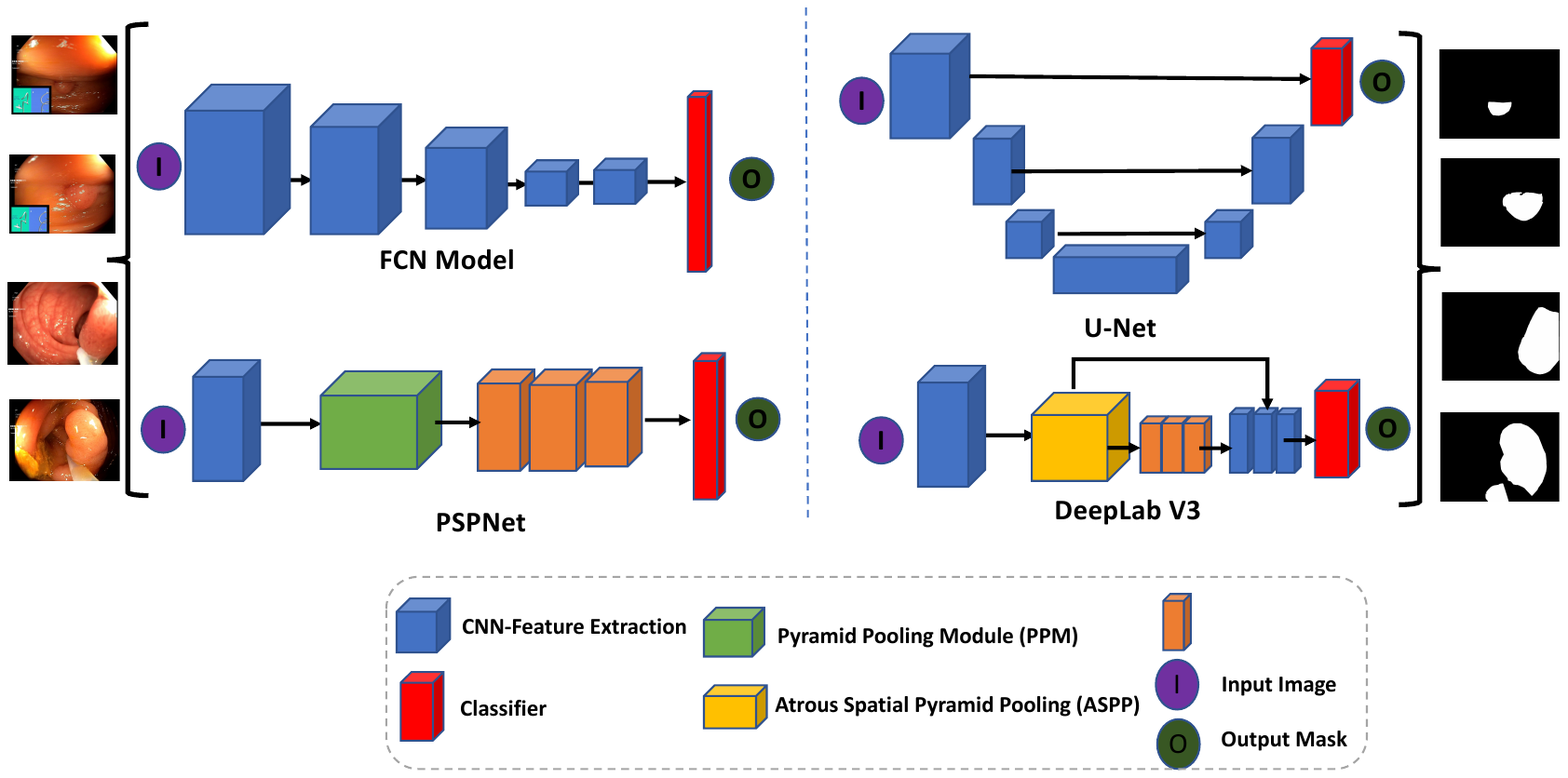}
    \caption{\textbf{Architecture layout for state-of-the-art methods}: FCN (fully convolutional Network), U-Net, PSPNet (Pyramid Scene Parsing Network) and DeepLabV3.}
    \label{SOTAIMG}
\end{figure}

{{Long et al. \cite{long2015fully} presented a Fully Convolutional Network (\textbf{FCN}) that uses downsampling and upsampling for image segmentation. The model is dived into two sectors, the first sector is responsible for extracting detailed feature maps through downsampling the spatial resolution of the image. The second sector is responsible for retrieving the location information through upsampling}}. The \textbf{U-Net} architecture developed by Ronnerberger et al.~\cite{ronneberger2015u} has shown tremendous success in medical image segmentation~\cite{sevastopolsky2017optic} including endoscopy~\cite{ali_objective_2020,ali2021_endoCV2020}.
U-Net is generally an encoder network followed by a decoder network, where,  convolution blocks followed by max-pooling downsampling are applied to the image to encode feature presentations at different multiple levels. Afterwards, the decoder projects semantically the distinguishing characteristics learnt by the encoder. The decoder is composed of upsampling and concatenation followed by a standard convolution function. The skip connections between downsampling and upsampling paths in the U-Net (i.e. which makes it symmetric) is the main difference between the U-Net and the FCN~\cite{ozturk2020comparison}. Pyramid Scene Parsing Network (\textbf{PSPNet})~\cite{zhao2017pyramid} is designed to incorporate global context information for the task of scene parsing {{using context aggregation depending on different regions to take advantage of global context information. Both local and global findings provide a more reliable final prediction.}}. {{The pyramid pooling module and CNN backbone with dilated convolutions are both present in the PSPNet encoder.}}. Similarly, dilated convolutions enabled the construction of semantic segmentation networks to effectively control the size of the receptive field that was incorporated in an a family of very effective semantic segmentation architectures, collectively named DeepLab~\cite{chen2018encoder}. The DeepLabV3 capture multi-scale information by employing the atrous convolution at multiple rates in a cascade or parallel multi-scale context through spatial pyramid pooling. Moreover, ResUNet \cite{zhang2018road} incorporates the benefits of both the ResNet and U-Net which allowed the design of a network with fewer parameters and improved segmentation performance. Fig.~\ref{SOTAIMG} provides an illustrative figure for the architecture of SOTA methods as explained. 

All of these networks has been explored for polyp segmentation in literature~\cite{guo2020polyp,jha2021real,nguyen2020contour}. Here, we benchmark our dataset on these popular deep learning model architectures. Out-of-sample generalisation results for both single frame (Table~\ref{tab:singleSeg}) and sequence data (Table~\ref{tab:seqSeg}) has been included in our technical validation of the presented data.


%

%
%
\begin{table}[t!]
\centering
\centering
\small
\caption{{Performance evaluation of SOTA segmentation methods on 88 single frames from centre 6 in an out-of-sample generalisation task. Top two values are presented in bold.}~\label{tab:singleSeg}}
\begin{tabular}{l|lllllllllc}
\hline
\textbf{Method} & \textbf{JI}  $\uparrow$               & \textbf{DSC} $\uparrow$               & \textbf{F2}   $\uparrow$              & \textbf{PPV}  $\uparrow$              & \textbf{Recall} $\uparrow$            & \textbf{Acc.} $\uparrow$             & \textbf{$\mathbf{d}_{AHD}$} $\downarrow$   & \textbf{NSD}$\uparrow$ & \textbf{MASD} $\downarrow$   & \multicolumn{1}{c}{\textbf{FPS}$\uparrow$}  \\ 
\hline
FCN8~\cite{long2015fully}  & 
 \begin{tabular}[c]{@{}l@{}}0.68\\$\pm$\small{0.30}
 \end{tabular}  & 
  \begin{tabular}[c]{@{}l@{}}0.76\\$\pm$\small{0.30}
 \end{tabular} & 
   \begin{tabular}[c]{@{}l@{}}0.75\\$\pm$\small{0.31}
 \end{tabular}
 & 
\begin{tabular}[c]{@{}l@{}}0.90\\$\pm$\small{0.15}
 \end{tabular}
 & 
 \begin{tabular}[c]{@{}l@{}}0.74\\$\pm$\small{0.31}
 \end{tabular}
 & 
 0.97 & 10.69 &0.49 & 37.61 & 44 \\ \hline
\vspace{1mm}
U-Net~\cite{ronneberger2015u}                & 
 \begin{tabular}[c]{@{}l@{}}0.55\\$\pm$\small{0.34}
 \end{tabular}
 & 
  \begin{tabular}[c]{@{}l@{}}0.63\\$\pm$\small{0.36}
 \end{tabular}
 & 
   \begin{tabular}[c]{@{}l@{}}0.64\\$\pm$\small{0.36}
 \end{tabular}
 & 
\begin{tabular}[c]{@{}l@{}}0.76\\$\pm$\small{0.31}
 \end{tabular}
& 
\begin{tabular}[c]{@{}l@{}}0.66\\$\pm$\small{0.37}
 \end{tabular}
 & 0.96 & 13.89 &0.45  & 93.02  & 21     \\
\hline
\vspace{1mm}
PSPNet~\cite{zhao2017pyramid} & 
\begin{tabular}[c]{@{}l@{}}0.72\\$\pm$\small{0.27}
 \end{tabular}
 & 
 \begin{tabular}[c]{@{}l@{}}0.80\\$\pm$\small{0.26}
 \end{tabular}
 & 
  \begin{tabular}[c]{@{}l@{}}0.79\\$\pm$\small{0.27}
 \end{tabular}
 & 
   \begin{tabular}[c]{@{}l@{}}0.88\\$\pm$\small{0.20}
 \end{tabular}
 & 
    \begin{tabular}[c]{@{}l@{}}0.79\\$\pm$\small{0.28}
 \end{tabular}
 & \textbf{0.98} & 10.39 &0.56  &34.12  & 31 \\
\hline
\begin{tabular}[c]{@{}l@{}}DeepLabV3+~\cite{chen2018encoder}\\(ResNet50) \end{tabular}
 & 
\begin{tabular}[c]{@{}l@{}}\textbf{0.75}\\$\pm$\small{0.28}
 \end{tabular}
 & 
 \begin{tabular}[c]{@{}l@{}} \textbf{0.81}\\$\pm$\small{0.27}
 \end{tabular}
 & 
  \begin{tabular}[c]{@{}l@{}} \textbf{0.80}\\$\pm$\small{0.28}
 \end{tabular}
 & 
   \begin{tabular}[c]{@{}l@{}} \textbf{0.92}\\$\pm$\small{0.17}
 \end{tabular}
 & 
\begin{tabular}[c]{@{}l@{}} 
{0.79}\\$\pm$\small{0.29}
\end{tabular}
&\textbf{0.98} & \textbf{9.95}  &0.62  & 41.04 & 47 
 \\
 \hline
 \begin{tabular}[c]{@{}l@{}}ResNetUNet~\cite{he2016deep,ronneberger2015u}\\(ResNet34)
 \end{tabular}
& 
\begin{tabular}[c]{@{}l@{}} {0.73}\\$\pm$\small{0.29}
\end{tabular}
 & 
\begin{tabular}[c]{@{}l@{}} {0.79}\\$\pm$\small{0.29}
\end{tabular}
 & 
\begin{tabular}[c]{@{}l@{}} {0.77}\\$\pm$\small{0.29}
\end{tabular}
 & 
\begin{tabular}[c]{@{}l@{}} \textbf{0.92}\\$\pm$\small{0.20}
\end{tabular} 
 & 
\begin{tabular}[c]{@{}l@{}} {0.78}\\$\pm$\small{0.29}
\end{tabular} 
& \textbf{0.98} & 10.04& 0.59 & 35.83  &\textbf{87}  
\\
\hline
\begin{tabular}[c]{@{}l@{}}DeepLabV3+~\cite{chen2018encoder}\\(ResNet101) \end{tabular} & 
\begin{tabular}[c]{@{}l@{}} \textbf{0.75}\\$\pm$\small{0.28}
\end{tabular} 
 &
 \begin{tabular}[c]{@{}l@{}} \textbf{0.82}\\$\pm$\small{0.27}
\end{tabular} 
 &
 \begin{tabular}[c]{@{}l@{}} \textbf{0.80}\\$\pm$\small{0.27}
\end{tabular} 
 &
 \begin{tabular}[c]{@{}l@{}} \textbf{0.92}\\$\pm$\small{0.18}
\end{tabular} 
 & 
  \begin{tabular}[c]{@{}l@{}} \textbf{0.81}\\$\pm$\small{0.27}
\end{tabular} 
& 
\textbf{0.98}& \textbf{9.67} & \textbf{0.64}  &\textbf{23.29}  & 33
\\
\hline
 \begin{tabular}[c]{@{}l@{}}ResNetUNet~\cite{he2016deep,ronneberger2015u}\\(ResNet101)
 \end{tabular} & 
  \begin{tabular}[c]{@{}l@{}} {0.74}\\$\pm$\small{0.29}
\end{tabular}
& 
\begin{tabular}[c]{@{}l@{}} {0.80}\\$\pm$\small{0.28}
\end{tabular} 
 & 
 \begin{tabular}[c]{@{}l@{}} \textbf{0.80}\\$\pm$\small{0.28}
\end{tabular} &
\begin{tabular}[c]{@{}l@{}} {0.93}\\$\pm$\small{0.14}
\end{tabular} 
 & 
 \begin{tabular}[c]{@{}l@{}} \textbf{0.80}\\$\pm$\small{0.29}
\end{tabular} 
 & \textbf{0.98} & 10.10 &\textbf{0.63}  &\textbf{27.38}  & 40 
 \\
 \hline
  \bottomrule
\multicolumn{11}{l}{{\textbf{JI}: Jaccard index } \hspace{.1cm} {\textbf{DSC}: Dice coefficient} \hspace{.1cm} {\textbf{F2}: Fbeta-measure, with $\beta = 2$} \hspace{.05cm} {\textbf{PPV}: positive predictive value}} \\
\multicolumn{11}{l}{{\textbf{Acc.}: overall accuracy} \hspace{.05cm} {\textbf{$\mathbf{d}_{AHD}$}: avg. Hausdorff distance} \hspace{.1cm} {\textbf{FPS}: frames per second}} \hspace{.05cm} \\
\multicolumn{11}{l}{
{{$\uparrow$: best increasing} \hspace{.05cm} {$\downarrow$: best decreasing} \hspace{.05cm} {NSD: Normalised Surface Dice} \hspace{.05cm} {MASD: Mean Average Surface Distance} }}
\\
\end{tabular}
\end{table}
\begin{table}[t!h!]
\caption{{Performance evaluation of SOTA segmentation methods on 432 sequence frames from centre 6 in an out-of-sample generalisation task. Top two methods are presented in bold.}~\label{tab:seqSeg}}
\centering
\centering
\small
\begin{tabular}{l|lllllllll}
\hline
\textbf{Method} & \textbf{JI}  $\uparrow$               & \textbf{DSC}  $\uparrow$           & \textbf{F2} $\uparrow$             & \textbf{PPV} $\uparrow$           & \textbf{Recall} $\uparrow$         & \textbf{Acc.} $\uparrow$            & \textbf{$\mathbf{d}_{AHD}$} $\downarrow$  & \textbf{NSD} $\uparrow$  & \textbf{MASD} $\downarrow$     \\ 
\hline
FCN8~\cite{long2015fully}                   
&
 \begin{tabular}[c]{@{}l@{}}0.56\\$\pm$\small{0.37}
 \end{tabular} 
& 
\begin{tabular}[c]{@{}l@{}}0.62\\$\pm$\small{0.38}
 \end{tabular} 
 & 
\begin{tabular}[c]{@{}l@{}}0.59\\$\pm$\small{0.37}
\end{tabular} 
 & 
 \begin{tabular}[c]{@{}l@{}}0.88\\$\pm$\small{0.28}
\end{tabular} 
 & 
\begin{tabular}[c]{@{}l@{}}0.63\\$\pm$\small{0.36}
\end{tabular} 
 & 0.95 & 9.84  & 0.43 & 32.97 \\
UNet~\cite{ronneberger2015u}   & 
\begin{tabular}[c]{@{}l@{}}0.43\\$\pm$\small{0.37}
\end{tabular} 
 & 
\begin{tabular}[c]{@{}l@{}}0.50\\$\pm$\small{0.39}
\end{tabular} 
 & 
\begin{tabular}[c]{@{}l@{}}0.47\\$\pm$\small{0.39}
\end{tabular} 
 & 
\begin{tabular}[c]{@{}l@{}}0.68\\$\pm$\small{0.41}
\end{tabular} 
 & 
\begin{tabular}[c]{@{}l@{}}0.62\\$\pm$\small{0.37}
\end{tabular} 
 & 0.95 & 11.22 &0.39 &   48.66   \\
PSPNet~\cite{zhao2017pyramid}   & 
\begin{tabular}[c]{@{}l@{}}0.58\\$\pm$\small{0.38}
\end{tabular} 
 & 
\begin{tabular}[c]{@{}l@{}}0.64\\$\pm$\small{0.38}
\end{tabular} 
& 
\begin{tabular}[c]{@{}l@{}}0.61\\$\pm$\small{0.38}
\end{tabular} 
& 
\begin{tabular}[c]{@{}l@{}}0.84\\$\pm$\small{0.33}
\end{tabular} 
& 
\begin{tabular}[c]{@{}l@{}}0.68\\$\pm$\small{0.35}
\end{tabular}
 & 0.96 & 9.84 & 0.49& 33.46 \\
\begin{tabular}[c]{@{}l@{}}DeepLabV3+~\cite{chen2018encoder}\\(ResNet50) \end{tabular}                                         & 
\begin{tabular}[c]{@{}l@{}}0.60\\$\pm$\small{0.37}
\end{tabular}
 & 
\begin{tabular}[c]{@{}l@{}}0.67\\$\pm$\small{0.37}
\end{tabular}
&
\begin{tabular}[c]{@{}l@{}}0.64\\$\pm$\small{0.37}
\end{tabular}
&
\begin{tabular}[c]{@{}l@{}}0.85\\$\pm$\small{0.31}
\end{tabular}
& 
\begin{tabular}[c]{@{}l@{}}\textbf{0.71}\\$\pm$\small{0.33}
\end{tabular}
& 0.96 & 9.63  & 0.51& 28.19 \\
 \begin{tabular}[c]{@{}l@{}}ResNetUNet~\cite{he2016deep,ronneberger2015u}\\(ResNet34)
 \end{tabular}                                          & 
\begin{tabular}[c]{@{}l@{}}0.59\\$\pm$\small{0.37}
\end{tabular}
& 
\begin{tabular}[c]{@{}l@{}}0.66\\$\pm$\small{0.38}
\end{tabular}
& 
\begin{tabular}[c]{@{}l@{}}0.63\\$\pm$\small{0.38}
\end{tabular}
& 
\begin{tabular}[c]{@{}l@{}}0.87\\$\pm$\small{0.30}
\end{tabular}
& 
\begin{tabular}[c]{@{}l@{}}0.70\\$\pm$\small{0.35}
\end{tabular}
 & 0.96 & 9.78 & 0.50& 27.10   \\
  \begin{tabular}[c]{@{}l@{}}DeepLabV3+~\cite{chen2018encoder}\\ResNet101\end{tabular} & 
\begin{tabular}[c]{@{}l@{}}\textbf{0.65}\\$\pm$\small{0.37}
\end{tabular}
 &
\begin{tabular}[c]{@{}l@{}}\textbf{0.71}\\$\pm$\small{0.37}
\end{tabular}
& 
\begin{tabular}[c]{@{}l@{}}\textbf{0.68}\\$\pm$\small{0.37}
\end{tabular}
 & 
\begin{tabular}[c]{@{}l@{}}\textbf{0.90}\\$\pm$\small{0.26}
\end{tabular}
& 
\begin{tabular}[c]{@{}l@{}}\textbf{0.73}\\$\pm$\small{0.34}
\end{tabular}
& 
\textbf{0.97} & \textbf{9.08} & \textbf{0.57}& \textbf{18.59}  \\
 \begin{tabular}[c]{@{}l@{}}ResNetUNet~\cite{he2016deep,ronneberger2015u}\\(ResNet101)
 \end{tabular}                                        & 
 \begin{tabular}[c]{@{}l@{}}0\textbf{.65}\\$\pm$\small{0.37}
\end{tabular}
 & 
 \begin{tabular}[c]{@{}l@{}}\textbf{0.70}\\$\pm$\small{0.37}
\end{tabular}
& 
 \begin{tabular}[c]{@{}l@{}}\textbf{0.68}\\$\pm$\small{0.37}
\end{tabular}
& 
\begin{tabular}[c]{@{}l@{}}\textbf{0.92}\\$\pm$\small{0.22}
\end{tabular}
& 
\begin{tabular}[c]{@{}l@{}}\textbf{0.71}\\$\pm$\small{0.35}
\end{tabular}
 & \textbf{0.96} & \textbf{9.20}& \textbf{0.57}&  \textbf{22.70} \\
 \hline
 \bottomrule
\multicolumn{10}{l}{{\textbf{JI}: Jaccard index } \hspace{.1cm} {\textbf{DSC}: Dice coefficient} \hspace{.1cm} {\textbf{F2}: Fbeta-measure, with $\beta = 2$} \hspace{.1cm} {\textbf{PPV}: positive predictive value}} \\
\multicolumn{10}{l}{{\textbf{Acc.}: overall accuracy} \hspace{.1cm} {\textbf{$\mathbf{d}_{AHD}$}: Average Hausdorff distance}  \hspace{.05cm} {$\uparrow$: best increasing} \hspace{.05cm} {$\downarrow$: best decreasing}}\\
\multicolumn{10}{l}{{NSD: Normalised Surface Dice} \hspace{.05cm} {MASD: Mean Average Surface Distance}}
\end{tabular}
\end{table}

\begin{table}[t!h!]
\centering
\caption{{Centre-wise performance evaluation of best approach while training using four centres, validating on individual centre, and testing on out-of-sample generalisation task (centre C6).}\label{tab:crossvalresults}}
\centering
\small
\begin{tabular}{l|llllllllr}
\hline
\textbf{Method} &\textbf{Val.} & \textbf{JI}  $\uparrow$ & \textbf{DSC}  $\uparrow$           & \textbf{F2} $\uparrow$  & \textbf{PPV} $\uparrow$  & \textbf{Recall} $\uparrow$ & \textbf{Acc.} $\uparrow$ & \textbf{$\mathbf{d}_{AHD}$} $\downarrow$     \\ 
\hline

\begin{tabular}[c]{@{}l@{}}DeepLabV3+~\cite{chen2018encoder}\\(ResNet50) \end{tabular} &C1 &  0.70$\pm$\small{0.32} & 0.76$\pm$\small{0.33} & 0.75$\pm$\small{0.33} & 0.85$\pm$\small{0.28} & 0.79$\pm$\small{0.30} & 0.98 & 4.77  \\

\begin{tabular}[c]{@{}l@{}}DeepLabV3+~\cite{chen2018encoder}\\(ResNet50) \end{tabular} &C2 & 0.72$\pm$\small{0.28} &0.79$\pm$\small{0.27} &0.78$\pm$\small{0.27} & 0.88$\pm$\small{0.23} & 0.80$\pm$\small{0.25}  & 0.97 & 4.70\\    

\begin{tabular}[c]{@{}l@{}}DeepLabV3+~\cite{chen2018encoder}\\(ResNet50) \end{tabular} &C3 & 0.74$\pm$\small{0.28} & 0.80$\pm$\small{0.27} & 0.80$\pm$\small{0.26} & 0.86$\pm$\small{0.24} & 0.83$\pm$\small{0.25} & 0.97 & 4.77\\

\begin{tabular}[c]{@{}l@{}}DeepLabV3+~\cite{chen2018encoder}\\(ResNet50) \end{tabular} &C4 &0.76$\pm$\small{0.28} &0.82$\pm$\small{0.27} &0.81$\pm$\small{0.28} & 0.91$\pm$\small{0.17} & 0.81$\pm$\small{0.28} & 0.98 & 4.77 \\

\begin{tabular}[c]{@{}l@{}}DeepLabV3+~\cite{chen2018encoder}\\(ResNet50) \end{tabular} &C5 &0.73$\pm$\small{0.31} &0.78$\pm$\small{0.31} &0.77$\pm$\small{0.32} & 0.93$\pm$\small{0.16} & 0.78$\pm$\small{0.31} & 0.98 & 4.88 \\ 






 \hline
 \bottomrule
\multicolumn{9}{l}{{Val.: Validation centre} \hspace{.05cm} {\textbf{JI}: Jaccard index} \hspace{.05cm} {\textbf{DSC}: Dice coefficient} \hspace{.05cm} {\textbf{F2}: Fbeta-measure, with $\beta = 2$} \hspace{.05cm} {\textbf{PPV}: Positive }} \\
\multicolumn{9}{l}{predictive value\hspace{.1cm}{\textbf{Acc.}: overall accuracy} \hspace{.1cm} {\textbf{$\mathbf{d}_{AHD}$}: Average Hausdorff distance} \hspace{.05cm} {$\uparrow$: best increasing} \hspace{.05cm} {$\downarrow$: best decreasing}}
\end{tabular}
\end{table}

\subsection*{Validation Summary}
Our technical validation suggest that DeepLabV3+ with ResNet101 has the best performance on most metrics except for FPS suggesting larger latency in inference (Table~\ref{tab:singleSeg} and Table~\ref{tab:seqSeg}). The highest score was 0.82 for DSC and the least 9.67 for $d_{AHD}$ with DeepLabV3+ with ResNet101 on single frame data. However, the second best inference speed (FPS of 47) and score (DSC $= 0.81$, $d_{AHD} = 9.95$) was obtained again using DeepLabV3+ but with ResNet50 backbone. Similarly, for sequence C6 out-of-sample test generalisation the highest score of 0.65 for DSC with highest recall of 0.73 and the least 9.08 for $d_{AHD}$ was obtained with DeepLabV3+ with ResNet101 backbone. With the same ResNet101 backbone ResUNet resulted in very close performance of 0.65 DSC but with highest precision on 0.92 and $d_{AHD}$ of 9.20. However, even with the same ResNet101 backbone ResUNet (40 FPS) has better speed compared to DeepLabV3+ (33 FPS).
{In addition, we also evaluated using normalised surface dice (NSD) and mean average surface distance (MASD) both of which demonstrated similar performance trend for most methods. NSD was the lowest for the DeepLabV3+ and ResNetUNet with ResNet101 backbone, 0.64 and 0.63, respectively for single frames, and 0.57 each for sequence frames. The lowest MASD was reported for DeepLabV3+ with ResNet101 backbone with 23.29 and 18.59, respectively, for single and sequence frames. We also ran size-stratified DSC estimates for each algorithms. For medium and large polyps, the DSC score for majority of methods were not affected with DSC of 0.87 for large polyps and 0.84 for medium in the case of DeepLabV3+ with ResNet101 backbone. However, a steep decrease in DSC was observed for the small polyps with DSC value of only 0.46. Also for classical PSPNet and FCN8 networks, DSC difference was estimated to be over 0.10 between large and medium polyp sizes while for both ResNet-UNet and DeepLabV3+ had much smaller difference.}

%
\section*{Discussion}
%
{While for the single frames data DSC is above 0.80 for ResNetUNet and DeepLabV3+ using ResNet101 backbone, however, the same architectures only provided DSC around 0.70 on the sequence dataset. This can be primarily because of the larger number of frames in the sequence dataset (nearly 5 times) which if 432 versus only 88 single frames, and the heterogeneous image quality in the sequence data. Additionally, it is to be noted that while single frames have very clean polyp images, sequence can have different view points, sizes and quality and may or may not consist of polyps as in the real-world colonoscopy data. The boundary distances comparing with the estimated and ground truth mask boundaries using NSD, MASD and HD showed similar performance changes between different methods.}

{While most methods have close to 40 FPS performance, ResNetUNet with smaller backbone architecture ResNet34 provided real-time performance with 87 FPS but with DSC close to DeepLabV3+ with ResNet50 backbone. However, when evaluated on boundary distance-based metric MASD, it can be observed (Table~\ref{tab:singleSeg}--Table~\ref{tab:seqSeg}) that the ResNetUNet have desired lower values in both single (35.83 vs 41.04) and sequence (27.10 vs 28.19) data compared to the DeepLabV3+. It can be inferred that MASD better captures the performance on small and medium sized polyps compared to other metrics.}

{The quantitative performances on this diverse multicentre dataset illustrates that utilising only baseline architectures such as vanilla UNet, PSPNet and FCN8 only provide sub-optimal results. This can be due to their model architectures that does not allow for capturing diverse polyp size and polyp variability and other changes such as color and contrast in images. However, using residual networks, atrous spatial pyramid pooling layers, and use of deeper backbones can improve generalisability of methods resulting in better performance on unseen centres. While small-sized polyp are generally ($\leq 100 \times 100$ pixels) performed poorly, these methods were able to capture the variabilities in both medium and large sized polyps. Additionally, as illustrated from our cross-validation results in Table~\ref{tab:crossvalresults} that even using networks that provide optimal results in this dataset, the choice of validation data affects the network performance largely.}

\begin{figure}[t!h!]
    \centering
    \includegraphics[width=0.9\textwidth]{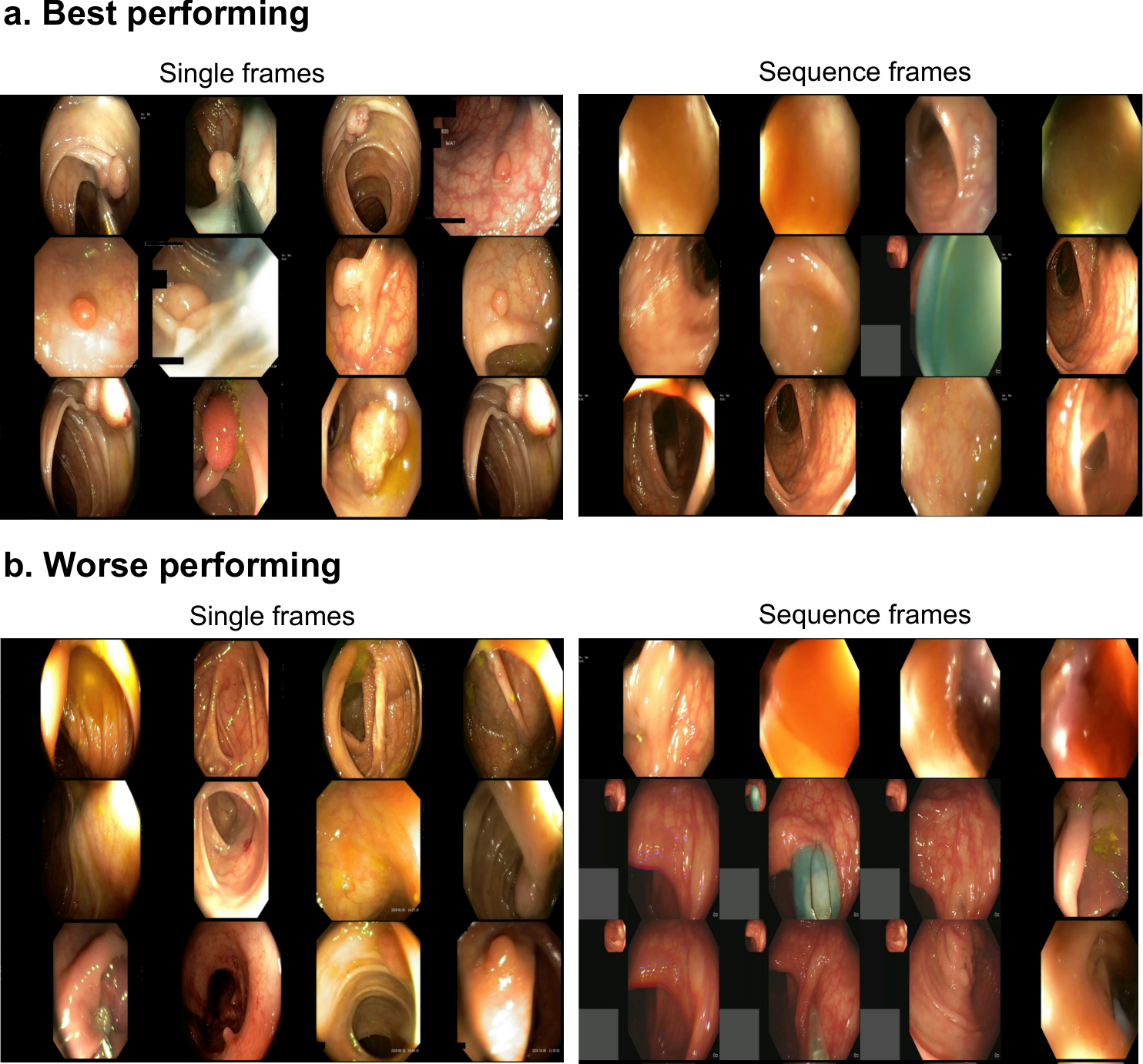}
    \caption{\textbf{Best and worse samples for single and sequence frames.} 12 images for top performing frames (a) with highest DSC scores and bottom performing frames (b) with lowest DSC.}
    \label{fig:top_bottom}
\end{figure}

{From the qualitative results presented in Fig.~\ref{fig:top_bottom}, it can be concluded that the best performing frames in the single frames are those which have clear polyp views with appearances that is either very different to background or at least appear as a lifted mucosa distinctly from the background mucosa. However, for the worse performing frames in the same, these appear to be most frames with either local strong illumination that confuses the network or embedded in the mucosa background color, i.e., almost flat or less protruded. Similarly, for the sequence frames, most frames that had good score were mostly with no polyp in it, while the networks failed to detect sessile or flat polyps.}

\section*{Limitations of the dataset}
The positive sample catalogue consists of both polyp and non-polyp images for completeness (see Figure~\ref{fig:polypGenDatasetOverview} b). {However, it has been made sure that these mimic the real-world dataset and taken such that non-polyp images are in close proximity to at least one polyp region.}
The entire dataset has been carefully reviewed by senior gastroenterologists. The accuracy, reliability and completeness of the annotations are subjective to the annotators.
One additional limitation of this dataset is that the ambiguous annotations were mostly removed. For future versions, we will aim to quantify the level of disagreement among experts for each frame instead.

\section*{Usage Notes}
All released dataset has been published under Creative Commons CC-BY licence. Dataset has been released only for educational, research and commercial purpose. Anyone using this dataset for their research or commercial application need to adhere to CC-by (Credit must be given to the creator) by citing this paper and acknowledging them.

The released dataset has been divided into positive and negative samples. Additionally, positive samples are divided into single frames and sequence frames. Users are free to use the samples according to their method demand. For example, for fully convolutional neural networks we adhere to the use of positive samples as done in our technical validation, while for the recurrent techniques that exploit temporal information users may use both positive and negative sequence data. 
\section*{Code availability}
To help users with the evaluate the generalizability of detection and segmentation method a code is available at: \url{https://github.com/sharibox/EndoCV2021-polyp_det_seg_gen}. The code also consists of inference codes that to assist in centre-based split analysis. Benchmark codes of the polypGen dataset with provided training and validation split in this paper for segmentation is also available at: \url{https://github.com/sharibox/PolypGen-Benchmark.git}.
All the method codes are also available at different GitHub repositories provided in the Table 1.

\bibliography{scienceData}

\section*{Acknowledgements}
The research was supported by the National Institute for Health Research (NIHR) Oxford Biomedical Research centre (BRC). The views expressed are those of the authors and not necessarily those of the NHS, the NIHR or the Department of Health. JE. East is supported by NIHR Oxford BRC. D. Jha was funded by PRIVATON project and J. Rittscher by Ludwig Institute for Cancer Research and EPSRC Seebibyte Programme Grant.  
%
\section*{Author contributions statement}
S. Ali conceptualized, initiated, and coordinated the work. He led the data collection, curation, and annotation processes and conducted most of the analyses and writing of the paper. T. de Lange assisted in writing of the introduction, clinical correctness of the paper and provided feedback regarding description of sequences presented in the manuscript. D. Jha and N. Ghatwary assisted in data annotation and parts of technical validation. S. Realdon, R. Cannizzaro, O. Salem, D. Lamarque, C. Daul, T. de Lange, M. Riegler, P. Halvorsen, K. Anonsen, J. Rittscher, and J. East were involved directly or indirectly in facilitating the video and image data from their respective centres. Senior gastroenterologists and collaborators S. Realdon, R. Cannizzaro, O. Salem, D. Lamarque, T. de Lange, and J. East provided timely review of the annotations and required feedback during dataset preparation. All authors read the manuscript, provided substantial feedback, and agreed for submission.
\section*{Competing interests}
J. E. East has served on clinical advisory board for Lumendi, Boston Scientific and Paion; Clinical advisory board and ownership, Satisfai Health; Speaker fees, Falk. A. Petlund is the CEO and T. de Lange serves as chief medical scientist at Augere Medical, Oslo, Norway. All other authors declare no known competing financial interests or personal relationships that could have appeared to influence the work reported in this paper.

\end{document}